\documentclass[%
 reprint,
superscriptaddress,
 amsmath,amssymb,
prb,
]{revtex4-2}

\usepackage{graphicx}
\usepackage{dcolumn}
\usepackage{bm}
\usepackage{hyperref}
\usepackage[group-separator={,}]{siunitx}   
\usepackage{textcomp}
\usepackage{amsmath}

\begin{document}

\preprint{APS/123-QED}

\title{Ultrafast behavior of induced and intrinsic magnetic moments in CoFeB/Pt bilayers probed by element-specific measurements in the extreme ultraviolet spectral range}

\author{Clemens von Korff Schmising}
 \email{korff@mbi-berlin.de}
 \author{Somnath Jana}
 \author{Kelvin Yao}
 \author{Martin Hennecke}
 \author{Philippe Scheid}
 \author{Sangeeta Sharma}
   \affiliation{%
 Max-Born-Institut für Nichtlineare Optik und Kurzzeitspektroskopie, Max-Born-Straße 2A, 12489 Berlin, Germany
}%
 \author{Michel Viret}
 \author{Jean-Yves Chauleau}
  \affiliation{%
SPEC, CEA, CNRS, Université Paris-Saclay, CEA Saclay - 91191 Gif sur Yvette, France
}%
 \author{Daniel Schick}
 \affiliation{%
 Max-Born-Institut für Nichtlineare Optik und Kurzzeitspektroskopie, Max-Born-Straße 2A, 12489 Berlin, Germany
}%
 \author{Stefan Eisebitt}
\affiliation{%
 Max-Born-Institut für Nichtlineare Optik und Kurzzeitspektroskopie, Max-Born-Straße 2A, 12489 Berlin, Germany
}%
\affiliation{Technische Universität Berlin, Institut für Optik und Atomare Physik, 10623 Berlin, Germany}

\date{\today}

\begin{abstract}
The ultrafast and element-specific response of magnetic systems containing ferromagnetic 3\textit{d} transition metals and 4\textit{d}/5\textit{d} heavy metals is of interest both from a fundamental as well as an applied research perspective. However, to date no consensus about the main microscopic processes describing the interplay between intrinsic 3\textit{d} and induced 4\textit{d}/5\textit{d} magnetic moments upon femtosecond laser excitation exist. In this work, we study the ultrafast response of CoFeB/Pt bilayers by probing element-specific, core-to-valence band transitions in the extreme ultraviolet spectral range using high harmonic radiation. We show that the combination of magnetic scattering simulations and analysis of the energy- and time-dependent magnetic asymmetries allows to accurately disentangle the element-specific response in spite of overlapping Co and Fe M$_{2,3}$ as well as Pt O$_{2,3}$ and N$_7$ resonances. We find a considerably smaller demagnetization time constant as well as much larger demagnetization amplitudes of the induced moment of Pt compared to the intrinsic moment of CoFeB. Our results are in agreement with enhanced spin-flip probabilities due to the high spin-orbit coupling localized at the heavy metal Pt, as well as with the recently formulated hypothesis that a laser generated, incoherent magnon population within the ferromagnetic film leads to an overproportional reduction of the induced magnetic moment of Pt.

\end{abstract}

\maketitle


\section{Introduction}

Combining ferromagnetic 3\textit{d} transition metals with 4\textit{d}/5\textit{d} heavy metals leads to magnetic systems with new macroscopic functionalities. Co/Pt multilayers or FePt nanoparticles, for example, exhibit very high magnetocrystalline anisotropies and are therefore important model systems for data storage technology.  Pt underlayers can be exploited for spin-orbit torque induced switching \cite{Brataas2012,Dieny2020} with potential for ultrafast applications \cite{Jhuria2020}.  Light-driven processes allowing the manipulation and control of ferromagnetic order are receiving renewed attention due to the discovery of helicity-dependent, all-optical switching in thin Co/Pt multilayers and FePt based granular films \cite{Lambert2014}, with promising new developments based on tailored double-pulse excitation schemes \cite{Yamada2022}.

The ultrafast response of optically excited ferromagnetic transition/heavy metal systems is characterized by their very different atomic spin-orbit coupling strength, their distinct electronic structures and importantly by the behavior of intrinsic 3\textit{d} versus induced 4\textit{d}/5\textit{d} magnetic moments. The phenomenon of proximity-induced magnetism is caused by hybridization of the 3\textit{d} and 4\textit{d}/5\textit{d} bands and leads to parallel spin alignment between intrinsic and induced moment \cite{Hellman2017, Nakajima1998}. The complex interplay of the involved elements after optical excitation has been studied in a growing number of element-specific experiments based on resonant X-ray or extreme ultraviolet spectroscopy, leading, however, to conflicting observations and corresponding competing theoretical explanations. While experiments with Co/Pt bi- \cite{Willems2015} or multi- \cite{Hennes2022} layers as well as FePt alloys \cite{Hofherr2018} suggest that the induced Pt moment follows the dynamics of the ferromagnetic transition metal, later work found clear evidence for a distinct, element-specific response: both, in an ordered FePt compound \cite{Yamamoto2019} as well as in a Co/Pt multilayer \cite{Yamamoto2020}. There, a significantly slower dynamics of Pt was found and rationalized with a higher mobility of Co or Fe compared to Pt majority electrons leading to an enhanced demagnetization rate of the transition metal due to superdiffusive spin currents. Additionally, ground-state density-of-state calculations have predicted that a potential generation of incoherent magnons would lead to an overproportional reduction of the induced compared to the intrinsic magnetic moments: as canting of the 3\textit{d} spins changes the average spin alignment between neighboring atoms, the exchange interactions on the 4\textit{d}/5\textit{d} heavy metal is reduced, leading to a reduction of the induced moment. These calculations showed that if the response of the transition metal is dominated by Heisenberg-like, transversal excitation, the induced moment behaves differently and exhibits a strong reduction of its \textit{amplitude} \cite{Yamamoto2019,Vaskivskyi2021}. The aforementioned study investigating a CoPt alloy, qualitatively confirmed this hypothesis,  revealing slightly larger demagnetization amplitudes of Pt compared to Co \cite{Vaskivskyi2021}. An explanation based on enhanced spin-orbit coupling of the heavy metal Pd and a respective stronger spin-flip probability was invoked to rationalize an accelerated demagnetization rate with increasing Pd concentrations in NiPd alloys \cite{Gang2018}. Finally, we suggested a scenario where optical intersite spin transfer (OISTR) \cite{Dewhurst2018b}, i.e., a laser driven transfer of minority carriers from Pt to Co, competes with spin-orbit driven spin-flips locally enhanced on the Pt atom, resulting in comparable magnetization dynamics of Co and Pt. Experimentally, this hypothesis was tested in an experiment investigating a CoPt alloy using a combination of ultrafast magnetic circular dichroism and helicity-dependent transient absorption spectroscopy in the XUV spectral range \cite{Willems2020a}. Also based on OISTR, ab-initio calculations of the ultrafast magnetization changes of a FePd$_3$ compound predicted element-specific dynamics with a qualitative dependence on the laser pulse intensity \cite{Elhanoty2022}.  

However, when evaluating and comparing these results in literature, it is important to acknowledge the different experimental techniques, different excitation levels and -- most importantly -- the different studied sample geometries, ranging from alloys with different stoichiometries to bi- and multilayers with different film thicknesses leading to very different local environments. While in reflection geometry with an enhanced surface sensitivity, super-diffusive spin currents may lead to a distinct 3\textit{d} versus 4\textit{d}/5\textit{d} response \cite{Hofherr2018,Yamamoto2019,Yamamoto2020,Vaskivskyi2021}, experiments in transmission geometry yield signals averaged over the entire sample volume and will be less affected by intralayer spin transport \cite{Willems2015,Willems2020a,Hennes2022}. OISTR is confined to spin transfer among nearest neighbors \cite{Dewhurst2018a,Chen2019} and is hence expected to be most pronounced in alloys with equal concentrations of Co and Pt or ultrathin layered systems.

In this study, we report on distinct, layer-dependent dynamics of a CoFeB(5.3 nm)/Pt(3.5 nm) bilayer observed by element-specific, transverse magneto-optical Kerr effect (T-MOKE) measurements in the XUV spectral range using radiation of a high-harmonic generation (HHG) source. We demonstrate that the combination of magnetic scattering simulations \cite{Schick2014,Schick2021c} and an analysis of the ultrafast response at different photon energies allows accurately separating the element-specific response of Co or Fe and Pt in spite of strongly overlapping resonances.  The experimental results reveal a transient magnetic state with a significantly faster and more efficient quenching of the induced Pt moments compared to the response of the transition metal CoFeB alloy.  A systematic comparison with a CoFeB/MgO/Pt system, exhibiting no induced magnetization of Pt, supports our analysis.

\section{Experiment}
\begin{figure}
\includegraphics[width=0.5\textwidth]{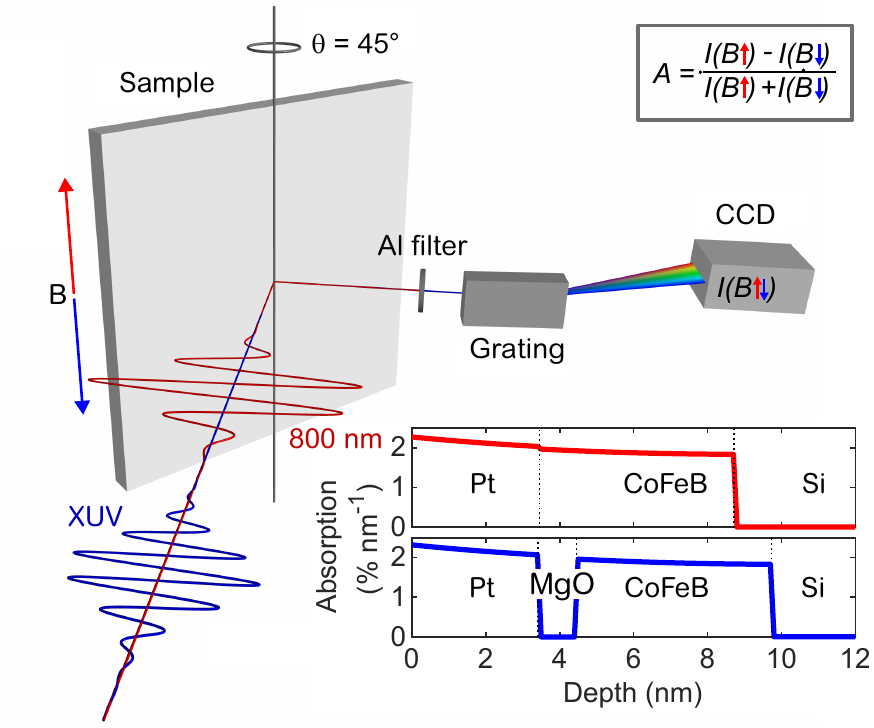}
\caption{Schematic of the T-MOKE geometry with an incident angle $\theta = 45^\circ$ of the \textit{p}-polarized XUV radiation. After reflection, the XUV pulses are spectrally dispersed by a flat field concave grating and detected by a CCD camera. An external magnetic field, $B$, is aligned perpendicularly to the incoming \textit{p}-polarized XUV light. The sample is excited by laser pulses at $\lambda = \SI{800}{nm}$ in a nearly collinear geometry. The inset shows the depth-dependent absorption profiles of the CoFeB/Pt and CoFeB/MgO/Pt samples. }
\label{fig:setup} 
\end{figure}
Si/Co$_{40}$Fe$_{40}$B$_{20}$(\SI{5.3}{nm})/MgO(\SI{1}{nm})/Pt(\SI{3.5}{nm}) as well as Si/Co$_{40}$Fe$_{40}$B$_{20}$(\SI{5.3}{nm})/Pt(\SI{3.5}{nm}) were deposited on Si substrates by magnetron sputtering. Both samples exhibit an in-plane magnetization and a square hysteresis loop with a small coercive field below \SI{2}{mT}. The magnetic films are protected against oxidation by the Pt layer or MgO/Pt bilayer. The thicknesses as well as interface roughnesses of below \SI{0.5}{nm} have been confirmed by independent X-ray reflectivity measurements. 

A schematic of the setup is shown in Fig.\ref{fig:setup}. Intense laser pulses at a repetition rate of \SI{3}{kHz}, a central wavelength of $\lambda = \SI{800}{nm}$ and a pulse length of \SI{25}{fs} are focused into a cell filled with neon gas, leading to higher harmonic radiation. The emitted XUV spectrum consists of discrete peaks with an approximate spectral width of \SI{200}{meV}. Additionally, we generate continuous spectra for the static characterization by averaging multiple acquisitions with varied chirp and peak intensities of the laser pulses as well as with varying gas pressures. The \textit{p}-polarized XUV radiation is incident on the sample at an angle of $\theta = 45^\circ$, reflected, and then spectrally dispersed by a flatfield concave grating and detected by a charged coupled device (CCD). The magnetization of the sample is set by applying an external magnetic field perpendicular to the incident polarization, facilitating the T-MOKE geometry \cite{La-O-Vorakiat2012}.  The magnetic asymmetry, $A$, is calculated as the normalized difference of two spectra, $I$, recorded for opposite magnetization directions of the CoFeB film, cf. Fig.\ref{fig:setup}. The time-resolved experiments were performed in a  pump-probe geometry using pump pulses with a central wavelength of \SI{800}{nm} and a pulse duration of $<\SI{30}{fs}$ as confirmed by autocorrelation measurements at the sample position. We estimate the upper bound for the temporal resolution of the experiment to be \SI{35}{fs}. The absolute values of incident fluences between 0 and \SI{12}{mJ\per\square\cm} are determined by a calibrated power meter and careful measurements of the beam area (full width at half maximum) using a beam profile camera. Calculations of the layer-dependent absorption profile based on a multilayer formalism \cite{Schick2014,Schick2021c}, using literature values of the optical refractive indices \cite{Choi2014,Hoffmann_2019}, show an almost homogeneous energy deposition witin the CoFeB and Pt layers (cf. Fig.\ref{fig:setup}).  To improve the signal-to-noise ratio, we acquire the incoming spectrum with a separate spectrometer for normalization and repeat each time-delay scan up to 50 times. For more details on the experimental setup please refer to Yao et al. \cite{Yao2020}.

\section{Results}
\begin{figure}
\includegraphics{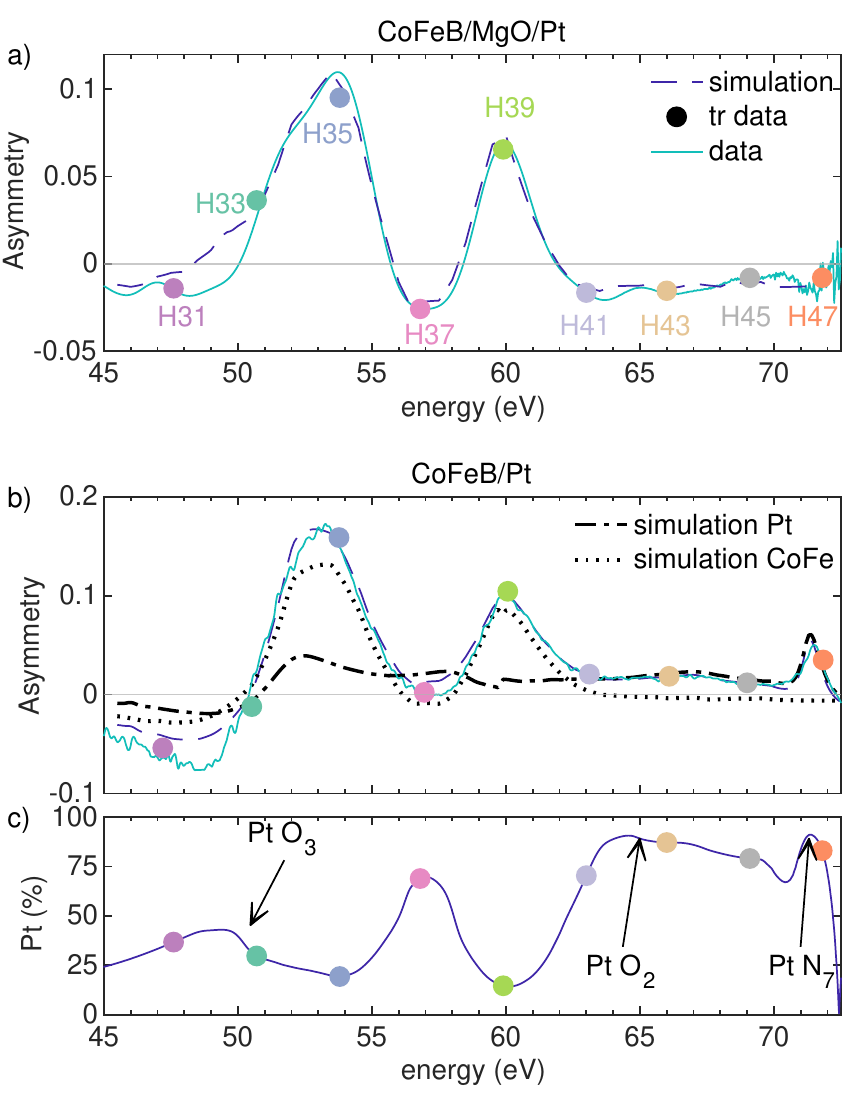}
\caption{Measured and calculated static magnetic asymmetry of a) CoFeB/MgO/Pt and b) CoFeB/Pt as a function of photon energy. The color-coded dots mark the photon energies of the time-resolved measurements, while the solid lines are static measurements with a continuous HHG spectrum. In b) we additionally show the calculated magnetic asymmetry stemming from CoFe (dotted line) and Pt (dashed line). The largest relative contributions to the magnetic asymmetry of Pt are not only found around the Pt O$_{2,3}$ and Pt N$_7$ edges, but also at \SI{57}{eV}, where contribution of Co and Fe approaches zero, cf. panel c).}
\label{fig:staticspectra} 
\end{figure}

In Fig.~\ref{fig:staticspectra}, we show the magnetic asymmetry in a photon energy range between 45~eV and 72.5~eV for both samples, a) CoFeB/MgO/Pt and b) CoFeB/Pt. We plot the asymmetry measured with a continuous HHG spectrum and mark the photon energies of the nine discrete HHG peaks (H31 to H47) used in the time-resolved experiments with colored dots. The simulations for static asymmetries are carried out with the udkm1Dsim toolbox \cite{Schick2014,Schick2021c} numerically calculating the polarization-dependent wave propagation taking into account refraction and reflections of the nanolayered sytems. We use published values of the atomic and magnetic form factors of Fe and Co \cite{Willems2019} and extract values for the induced moment of Pt from measurements of CoPt \cite{,Willems2015,Dewhurst2020a}. Atomic form factors of MgO and B are taken from Henke et al. \cite{Henke1993}.  We keep the thicknesses and interface roughnesses fixed, but allow small variations in the densities as well as in the absolute values of the elemental magnetic moments. For the CoFeB/Pt system, we fix the thickness of the magnetized Pt layer at the interface to \SI{1}{nm} \cite{Nakajima1998,Geissler2001, Suzuki2005}.  The fitting results yield a good agreement with the experimental measurements, for both samples, we can clearly note the dominant contributions from the Fe and Co M$_{2,3}$ edges around \SI{53.5}{eV} and \SI{60}{eV} probed by the HHG emission peaks H35 and H39. In the energy range between \SI{47}{eV} and \SI{51}{eV} both samples exhibit qualitatively similar deviations between simulation and experiment. As in the CoFeB/MgO/Pt sample Pt is not magnetic, we can rule out that they are caused by incorrect magnetic form factors of Pt. For CoFeB/Pt, the simulation also slightly overestimates the small magnetic asymmetries around \SI{57}{eV}, probed by HHG peak H37. For the CoFeB/Pt sample, we extract the contributions of Pt to the magnetic asymmetry by setting the magnetic moment of Co and Fe to zero. This is shown as a dashed-dotted line in Fig.\ref{fig:staticspectra}~b). The contribution of the CoFeB layer is calculated accordingly and shown as a dotted line. While this approach neglects the influence of the dichroic part of the refractive index of either Pt and Co or Fe to the reflectivities, we validate this approximation by finding an excellent agreement between the sum of the individual asymmetries and the total asymmetry of CoFeB/Pt.  In Fig.\ref{fig:staticspectra} c), we plot the relative contribution of Pt to the magnetic asymmetry of the CoFeB/Pt sample, which is determined by the complex spectral overlap of the three magnetic elements Co, Fe and Pt. The asymmetry of harmonic H35 is dominated by Fe (80\%) and H39 is dominated by Co (85\%). The maximum relative contribution of Pt is at the Pt N$_7$ edge at \SI{71.2}{eV} with close to 90\%. The small asymmetries beyond the Co edge (harmonic peaks H41 to H45) are also dominated by Pt and yield values $>80\%$. Furthermore, the harmonic peak H33 and H37 exhibit appreciable Pt O$_{2,3}$ contributions of 30$\%$ and 70$\%$.  The system CoFeB/MgO/Pt is not expected to exhibit a static induced magnetization of Pt, which  -- within our signal-to-noise ratio – is also consistent with our simulation. The subtle differences of the energy-dependent magnetic asymmetries between both sample systems highlights the importance of dedicated calculations in order to extract elements-specific information in the XUV spectral range. 

\begin{figure}[h!]
\includegraphics{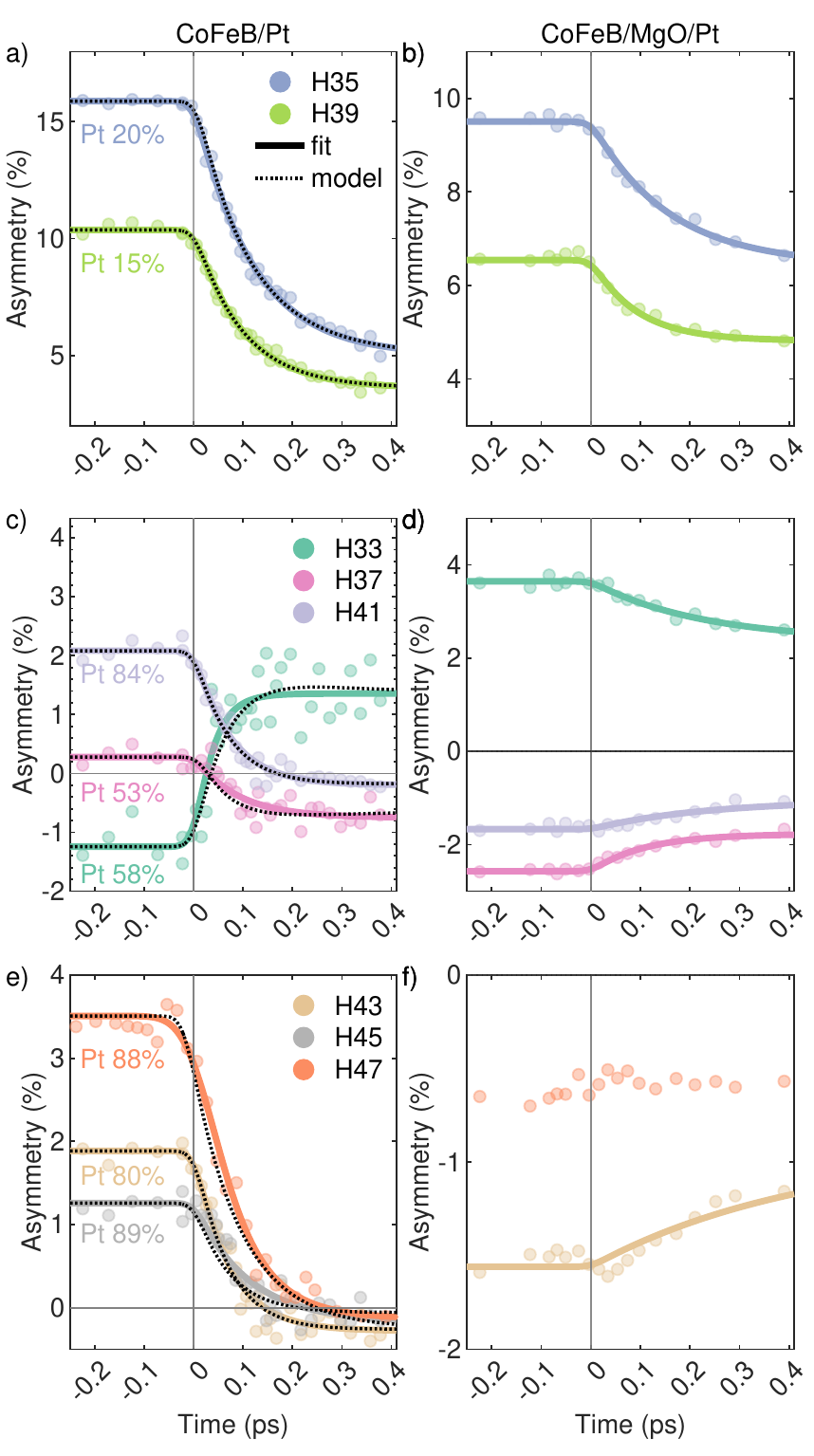}
\caption{Magnetic asymmetry as a function of time after optical excitation. Harmonics H35 and H39 in panel a) and b) are dominated by the dynamics of Fe and Co, respectively. For the CoFeB/Pt system, HHG peaks H33 and H37 (panel c) are sensitive to both layers, CoFeB and Pt and are characterized by a sign change of the time-resolved asymmetry. The asymmetries of the HHG peaks H41 (panel c), H43, H45 and H47 (panel a) are dominated by Pt contributions and -- differently to H35 and H39 -- exhibit a 100\% reduction of $A$.  For the CoFeB/MgO/Pt sample system all shown harmonics H33 to H47 show the same relative demagnetization amplitude and represent the response of Fe or Co. Harmonic H47, close to the Pt N$_7$ edge, shows no evidence of an increase, i.e. no evidence for tunneling of spin-polarized carriers from the CoFeB to Pt via the MgO layer, cf. panel f). }
\label{fig:time} 
\end{figure}

A summary of time-resolved data is shown in Fig.~\ref{fig:time} for very early times after optical excitation. We plot data for two exemplary fluences: \SI{10}{mJ\per\square\cm} and \SI{4}{mJ\per\square\cm} for the samples CoFeB/Pt and CoFeB/MgO/Pt, respectively. The solid lines are fits using a single-exponential decay convolved with a Gauss function, $G(t)$, reflecting the temporal resolution of the experiment:
\begin{equation}
    A(t) = G(t) \ast \left[A_0 + \Theta(t)\left( B \left(e^{-t/\tau} -1\right) \right)\right].
    \label{equ:fit}
\end{equation}
$\Theta(t)$ is the Heavyside function, $A_0$ is the static asymmetry and $B$ the amplitude of the demagnetization, $\tau$ denotes the demagnetization time constant. 
We differentiate the ultrafast response of the CoFeB/Pt sample measured at the different harmonics according to their element specificity:  H35 and H39 are dominated by Fe and Co, respectively and are shown in panel a), H33, H37 and H41 display a significant admixture with Pt an are summarized in panel c) and finally H43, H45 and H47 predominately reflect the ultrafast dynamics of Pt and are shown in panel e).  The response of the CoFeB/MgO/Pt sample at corresponding photon energies is shown in panels b), d) and f). 
For the CoFeB/Pt sample, we find demagnetization time constants for H35 $\tau_\mathrm{H35} = (115 \pm 7)$\si{fs} and for H39 $\tau_\mathrm{H39} = (105 \pm 6)$\si{fs}, for the CoFeB/MgO/Pt sample the values amount to $\tau_\mathrm{H35} = (140 \pm 20)$\si{fs} and $\tau_\mathrm{H39} = (100 \pm 19)$\si{fs}. Within our signal-to-noise ratio, we do not find any systematic variations of the demagnetization constants neither for the two different samples nor as a function of fluence. However, the slightly faster demagnetization constant of H39 (Co) compared to H35 (Fe) is consistently reproduced and is also in agreement with previous observations that $\tau$ scales with the magnetic moment of the element \cite{Koopmans2010, Radu2015}. The average of the extracted demagnetization time constants of harmonics H41 to H47 amounts to $\tau_{H41-H47} = (70\pm20)$\,fs, significantly smaller than what we find for H35 (Fe) and H39 (Co). Furthermore, the magnetic asymmetries measured for CoFeB/Pt at photon energies dominated by Pt, i.e. H41, H43, H45 and H47 shown in Fig.~\ref{fig:time} c) and e) are completely quenched or even show small negative values at $t=\SI{400}{fs}$.  This is in contrast to asymmetry values of harmonics H35 and H39, which are only reduced by ~35\% at this time delay.

\begin{figure}[ht!]
\includegraphics{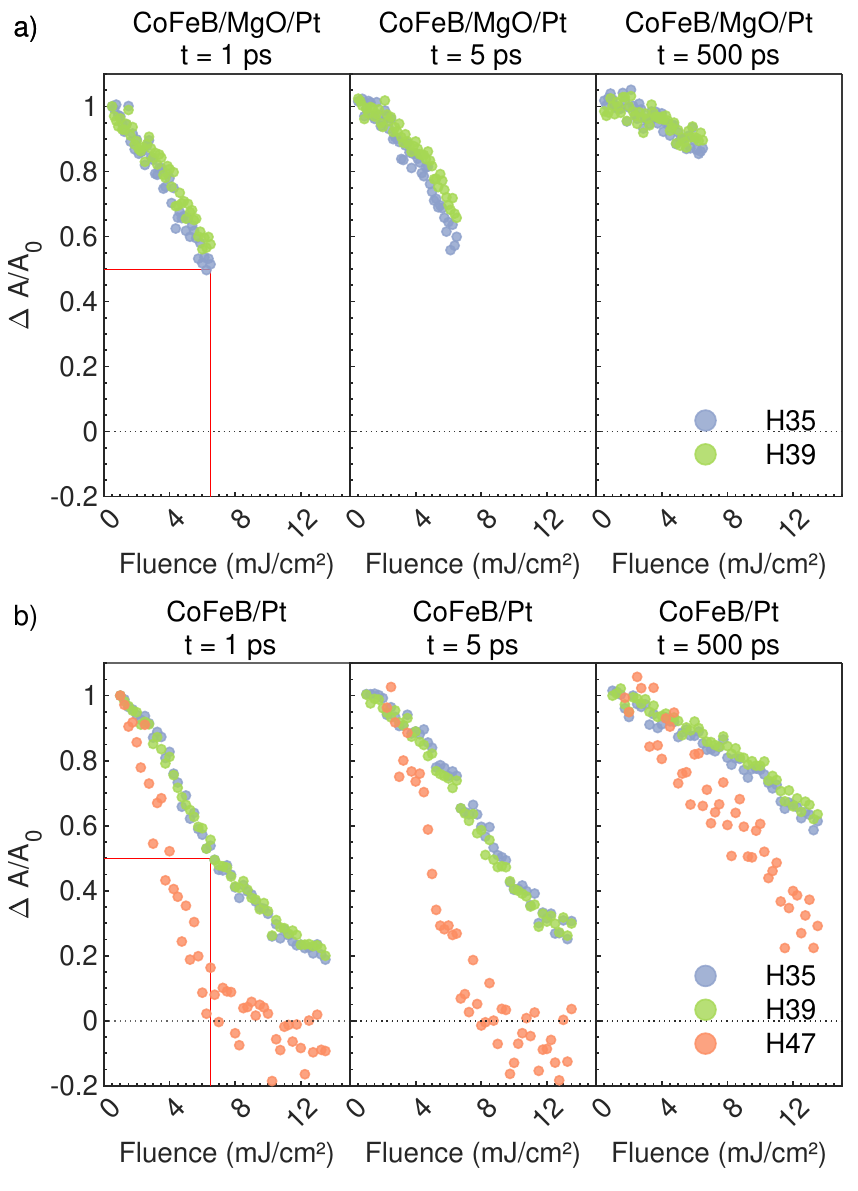}
\caption{Normalized asymmetry $\Delta A/A_0$ as a function of excitation fluence for three selected time delays, $t = 1, 5, 500$~\si{ps}. In panel a), we show H35 (Fe) and H39 (Co) for the CoFeB/MgO/Pt and in b) additionally H47 (Pt) for the CoFeB/Pt system. Pt demagnetizes significantly more efficiently and remains out of equilibrium with the CoFeB layer for up to \SI{500}{ps}.}
\label{fig:fluence} 
\end{figure}

The demagnetization amplitudes for different photon energies are more easily compared by inspection of Fig.~\ref{fig:fluence}, where we display the normalized magnetic asymmetries, $\widehat A = \Delta A/A_0$, as a function of the incident fluence at three selected time delays of $t = 1,5$ and 500~\si{ps}. Data for harmonics H35 and H39 show very similar values over the entire fluence range and for the three different times. This also applies for both sample systems: the red lines mark $\Delta A(t=1\mathrm{ps})/A_0 = 0.5$, which is reached for an incident fluence of \SI{6.5}{mJ\per\square\cm} for both samples. In Fig.~\ref{fig:fluence} b), we additionally show the magnetic asymmetry of harmonic H47 measured for sample CoFeB/Pt. Clearly, here $\Delta A/A_0$ is reduced more efficiently and already reaches zero for a fluence of $\approx$\SI{8}{mJ\per\square\cm}. Also note that even after a time delay of \SI{500}{ps}, when the CoFeB layer has recovered a significant fraction of its magnetization by energy dissipation, the magnetization of the Pt layer remains out of equilibrium with CoFeB. With the assignment of the different harmonic peaks to the different magnetic elements within the sample, this presents strong evidence that the interfacial Pt layer loses its induced magnetization significantly more efficiently than Co and Fe.

The element-specific dynamics allows us to use the time-resolved data to further verify and refine the relative elemental contributions to the magnetic asymmetry discussed in Fig.~\ref{fig:staticspectra} c). Accordingly, we describe the normalized nonlinear least square fits of the data (solid lines in Fig.~\ref{fig:time}) by a sum of equation \ref{equ:fit} representing separately the response of Co/Fe and Pt.
\begin{equation}
    \widehat A(t) = c_\mathrm{Pt} \widehat A_\mathrm{Pt}(t) + c_\mathrm{Co/Fe} \widehat A_\mathrm{Co/Fe}(t)\\
\label{equ:model}
\end{equation}
We use the extracted demagnetization time constants measured for H35, H39 as well as the average of H41 to H47 and assign them to Co, Fe and Pt, respectively. The normalized demagnetization amplitude, $\widehat B$, of Pt is assumed to be 100\%. We then describe our normalized, time dependent data, $\widehat A(t)$, by Eqn.\ref{equ:model} and only treat the demagnetization amplitude, $\widehat B$, of Co or Fe as well as the relative weights $c_\mathrm{Pt}$ and $ c_\mathrm{Co/Fe}$ as free parameters. Additionally the following condition has to hold: $c_\mathrm{Pt} + c_\mathrm{Co/Fe} = 1$.  The results are shown as dotted lines in Fig.~\ref{fig:time} a),c) and e). Analysis of the response of H35 and H39 yield demagnetization amplitudes for Fe and Co of $(40\pm5)\%$. The extracted relative Pt contributions are shown next to the time-resolved transients. We find a good agreement with the values determined from the static simulation, except for H33 and H37, where we see a substantial deviation. We attribute this to the very low values of the absolute asymmetry values in this energy range. 
A remarkable feature of the time traces of H33 and H37 is the sign change of the asymmetry. This is caused by the opposite sign of the respective asymmetries as well as the very different ultrafast response of the FeCoB and Pt layers. As Pt completely demagnetizes and CoFeB does not, the asymmetries reverse sign, which is quantitatively reproduced by Eq.\ref{equ:model}. The same argument holds for photon energies H43 to H47, where we also observe a zero crossing of $A$. Importantly, the observed sign change of the asymmetry is not related to a magnetization reversal.  Another potential explanation for the sign change of $A$ of H33 and H37 could be related to a spectral reshaping of $A$, as these photon energies probe the system very close to a zero crossing of the static asymmetry spectra. Such spectral reshaping, effectively shifting the resonant absorption and circular dichroism to smaller energies have been recently postulated for the XUV spectral range and are understood to be caused by laser driven changes of electron occupations \cite{Carva2009,Yao2020c,Rosner_2020}. However, as the slope of $A$ at energies H33 and H37 are very different, a laser induced spectral shift would result in significantly different dynamic changes, which is contrary to our observation.
Combining the static and time-resolved analysis, we estimate the uncertainty in assigning the relative elemental contribution for HHG peaks H33 and H37 of up to 15\%, for all other peaks to $<5\%$. 

A closer inspection of the ultrafast response of CoFeB/MgO/Pt in Fig.~\ref{fig:time} b),d),f) reveals comparable ultrafast dynamics with demagnetization amplitudes of $\Delta A/A_0 = (30\pm 5)\%$ at $t = \SI{0.4}{ps}$ for all HHG peaks.  The magnetic asymmetry measured at the HHG peak H47, probing the Pt N$_7$ edge, does not show any evidence of a magnetization increase, but stays constant within our signal-to-noise ratio. This implies that we can exclude significant ultrafast spin tunnelling through the MgO barrier into Pt, in agreement with a very recent study that found a exponential decay length of only \SI{2}{\angstrom} for the tunneling spin current through MgO  \cite{Wahada2022}. Just as for the CoFeB/Pt system, we observe similar demagnetization amplitudes of Fe and Co for longer timescales (cf. Fig.~\ref{fig:fluence} a), pointing towards an identical temperature dependence of the equilibrium magnetization of Co and Fe. This differs from previous observations on Co$_{60}$Fe$_{20}$B$_{20}$/SiO$_2$ systems, where a difference in the asymptotic limit of the remagnetization process was observed \cite{Hofherr2018}. 

\section{Discussion and Conclusions}

Our observations of distinct ultrafast dynamics of \textit{intrinsic} vs. \textit{induced} moments can be reconciled with our earlier work on Co$_{50}$Pt$_{50}$ alloys, where we identified two competing processes: spin-orbit induced spin flips and OISTR  \cite{Willems2019}: we argue again that the more efficient demagnetization of Pt may be a consequence of the significantly stronger spin-orbit coupling localized at the Pt atoms compared to Co or Fe \cite{Kuiper2014,Shanavas2014}. At the same time, with fewer Co or Fe-Pt nearest neighbors in the studied bilayer, competing OISTR processes are less likely to occur than in an alloy, leading to distinct dynamics between Co or Fe and Pt in this work.

Qualitatively, our results of a larger demagnetization amplitude of the Pt layer may also support the hypothesis that the induced moment of Pt is very sensitive to the orientation of the intrinsic CoFe moments and could hence be interpreted as an indicator for the generation of incoherent magnons in the CoFeB layer. In Yamamoto et al. \cite{Yamamoto2019}, however, it was postulated that magnons are predominantly generated after relaxation of the non-equilibrium electron distribution and will therefore only start to dominate after $\approx \SI{1}{ps}$, i.e. on a slower timescale compared to our observations. On the other hand, a number of recent studies have found evidence for an ultrafast generation of incoherent magnons upon laser excitation \cite{Carpene2015,Turgut2016,Eich2017,Gort2018,Zusin2018,Higley2019}.  Also, our observation that the magnetization of the Pt layers remains out of equilibrium with the adjacent CoFeB layer for delays larger than \SI{500}{ps}, i.e. much longer than heat equilibration times, may count as a further indication that induced moments of Pt are very sensitive to long-lived magnon populations of the polarizing Co or Fe moments. We anticipate that the combination of time-dependent density of state calculations with experiments reaching a significantly improved temporal resolution may be able to disentangle the different processes in the time domain and identify the time scale on which such a potential magnon population emerges. \\
Our results do not show any clear signatures of superdiffusive spin currents, in spite of also using a reflection geometry with depth-dependent sensitivity \cite{Hennecke2022}. A potential interlayer spin transport would be characterized by injection of majority carriers from CoFeB to the interface region of the Pt layer, leading to an increase of the demagnetiation rate of the CoFeB layer while simultaneously decreasing the demagnetization rate in the Pt layer, which is different to our observation. Also, the almost identical demagnetization rate of the CoFeB/Pt and CoFeB/MgO/Pt system (cf. red lines in Fig.~\ref{fig:fluence}) rules out significant interlayer superdiffusive spin currents, as transport would be inhibited in the latter sample by the insulating MgO layer.\\
In conclusion, we have presented a systematic comparison of two sample systems with and without an induced magnetization of Pt, CoFeB/Pt and CoFeB/MgO/Pt, and demonstrated that a combination of calculated magnetic asymmetries taking into account the geometry of the sample structure and an analysis of time-resolved magnetization traces allows to accurately disentangle the element-specific magnetic response in spite of strongly overlapping Co or Fe M$_{2,3}$ and Pt O$_{2,3}$ as well as N$_7$ resonances. We find a significantly more efficient demagnetization rate of the induced moment of Pt compared to the intrinsic moment of the transition metals in a CoFeB/Pt bilayer. Our results are in agreement with the presence of enhanced spin-orbit driven spin flips localized on Pt as well as with the recently postulated hypothesis of a strong sensitivity of an induced 4\textit{d}/5\textit{d} magnetic moment to the potential ultrafast emergence of an incoherent magnon population.

\begin{acknowledgments}
C. v. K. S., S. S. and S. E. acknowledge financial support by the Deutsche Forschungsgemeinschaft (DFG, German Research
Foundation) – Project-ID 328545488 – TRR 227, project A02 and A04. J.-Y. C. acknowledges funding from the French National Research Agency ANR
JCJC through the SPINUP project ref. ANR-21-CE24-0026-01.
\end{acknowledgments}

\bibliography{references.bib}

\end{document}